\documentclass{elsart}

 \usepackage{graphics}
 \usepackage{graphicx}
\usepackage{color}
 \usepackage{epsfig}
\usepackage{amssymb}

\begin{document}
%\linenumbers

\begin{frontmatter}
%  \linenumbers



  \title{Capability of the PAMELA Time-Of-Flight to
    identify light nuclei: results from a beam test calibration }


  \author[infnnapoli]{D. Campana},
  \author[infnnapoli,roma2]{R. Carbone},
  \author[napoli]{G. De Rosa},
  \author[infnnapoli]{G. Osteria \corauthref{cor1}},
  \author[napoli]{S. Russo},
  \author[siegen]{W. Menn},
  \author[roma2]{V. Malvezzi},
  \author[roma2]{L. Marcelli},
  \author[roma2]{P. Picozza},
  \author[roma2]{R. Sparvoli},
  \author[firenze]{L. Bonechi},
  \author[infnfirenze]{M. Bongi},
  \author[infnfirenze]{S. Ricciarini},
  \author[infnfirenze]{E. Vannuccini}

  \address[infnnapoli]{INFN, Section of Naples, Naples (Italy)}
  \address[napoli]{Dept. of Physics, University of Naples and INFN, Naples (Italy)}
  \address[siegen]{Dept. of Physics, University of Siegen, Siegen (Germany)}
  \address[roma2]{Dept. of Physics, University of Rome Tor Vergata and INFN, Rome (Italy)}
  \address[firenze]{Dept. of Physics, University of Florence and INFN, Florence (Italy)}
  \address[infnfirenze]{INFN, Section of Florence, Florence (Italy)}

  \corauth[cor1]{Corresponding authour:
    Address: Complesso Universitario di Monte S.Angelo Via Cintia  80126 Naples - ITALY;
    Phone: +39 081 676167;
    E-mail: giuseppe.osteria@na.infn.it}

  \begin{abstract}
    PAMELA is a space telescope orbiting around the Earth since June 2006.
    The scientific objectives addressed by the mission are the measurement of
    the antiprotons and positrons spectra in cosmic rays, the hunt for anti-nuclei as
    well as the determination of light nuclei fluxes from Hydrogen to Oxygen in a wide
    energy range and with very high statistics.
    In this paper the charge discrimination capabilities of the PAMELA Time-Of-Flight system for
    light nuclei, determined during a beam test calibration, will be presented.
  \end{abstract}


\end{frontmatter}

% main text

%--------------------------------------------
%%%%%%%%%%%%%%%%%%%%%%%%%%%%%%%%%%%%%%%%%%%%%
%%%%%%%%%%%%%%%%%%%%%%%%%%%%%%%%%%%%%%%%%%%%%
\section{Introduction}
%%%%%%%%%%%%%%%%%%%%%%%%%%%%%%%%%%%%%%%%%%%%%
%%%%%%%%%%%%%%%%%%%%%%%%%%%%%%%%%%%%%%%%%%%%%
%--------------------------------------------
\label{intro}

The PAMELA apparatus is designed to study charged particles in the cosmic radiation.
It is hosted by a Russian Earth-observation satellite, the Resurs-DK1, that was launched into space
by a Soyuz  rocket on the 15$^{th}$ June 2006 from the Baikonur cosmodrome (Kazakhstan). The satellite orbit is elliptical
and semi-polar, with an inclination of 70.0$^\circ$ and an altitude varying between 350\,km and 600\,km.
The mission will last nominally for three years. The main scientific goal of the experiment is the precise
measurement of the cosmic-ray antiproton and positron energy spectra. The satellite orbit and the mechanical
design of the apparatus allow the identification of these particles in an unprecedented energy range
(between 50\,MeV and 270\,GeV for positrons and between 80\,MeV and 190\,GeV for antiprotons) and
with high statistics ($\sim$ 10$^4$ antiprotons and $\sim$ 10$^5$ positrons per year). Additionally
PAMELA is searching for antimatter in the cosmic radiation, with a sensitivity for the anti-He/He
ratio of the order of $\sim$ 10$^{-7}$.

PAMELA is also aimed to extensively study the abundances and composition of light cosmic rays (up
to Oxygen) over almost three decades of energy. In order to clarify the role of the different
mechanisms that act in the propagation and transport of galactic cosmic rays it is fundamental
to have more precise and extended data on the relative abundances of the constituents of galactic
cosmic rays and especially on the ratio of secondary to primary particles, such as the Boron/Carbon
ratio. 

This paper will illustrate the light-charge identification capabilities of the PAMELA
Time-Of-Flight (TOF) system, as evaluated during a beam test performed at the GSI laboratory - Germany - in
February 2006. The TOF system is a key detector for the PAMELA instrument, providing trigger for
acquisition, measuring the particle flight time (necessary to reject the albedo background component)
and determining the absolute value of the particle charge. The paper is organized as follows: after an overview
of the PAMELA instrument (section \ref{The_PAMELA_instrument}), the GSI beam test (set-up of the detector, available beams, operational details)
is described in section \ref{btest}. Section \ref{TOF_time_resolution} is dedicated to the determination
of the TOF time resolution from the beam data, obtained with two different methods, and finally section
\ref{Charge_resolution_for_light_nuclei} reports the charge resolution of PAMELA TOF system for
several elements, ranging from Proton to Carbon.

%--------------------------------------------
%%%%%%%%%%%%%%%%%%%%%%%%%%%%%%%%%%%%%%%%%%%%%
%%%%%%%%%%%%%%%%%%%%%%%%%%%%%%%%%%%%%%%%%%%%%
\section{The PAMELA instrument and the Time-Of-Flight system}
%%%%%%%%%%%%%%%%%%%%%%%%%%%%%%%%%%%%%%%%%%%%%
%%%%%%%%%%%%%%%%%%%%%%%%%%%%%%%%%%%%%%%%%%%%%
%--------------------------------------------
\label{The_PAMELA_instrument}

\begin{figure}
  \begin{center}
    \mbox{\includegraphics*[scale=0.4]{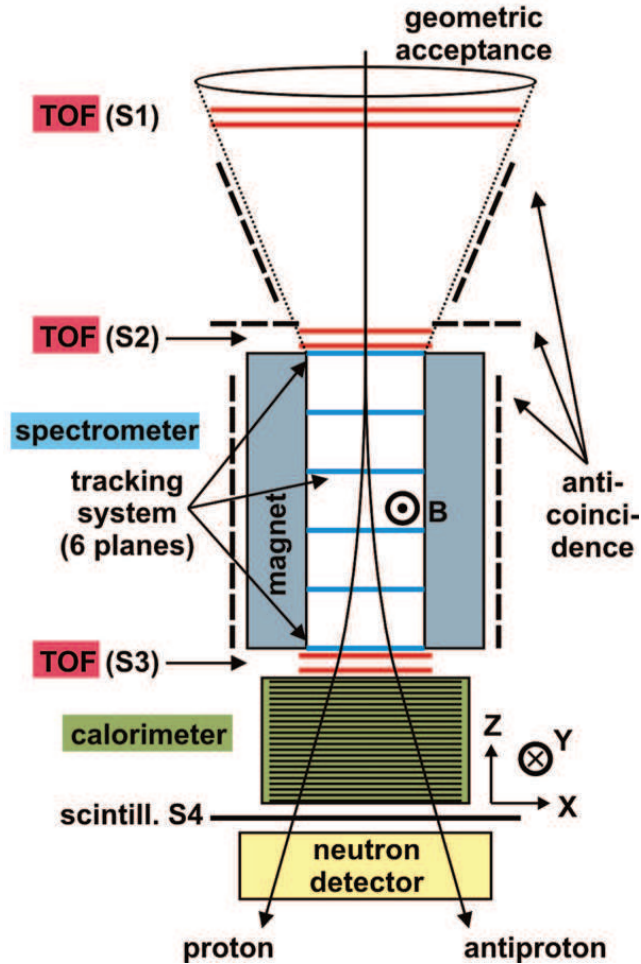}}
    %\hspace{-3mm}
    %\includegraphics*[height=7cm,width=0.5\textwidth,angle=0,clip]{pam.eps}
    \caption{A sketch of the PAMELA telescope. The method of discrimination between particle
      and antiparticle with the magnetic spectrometer is illustrated. The main direction of
      the magnetic field B inside the spectrometer is also shown.}
    \label{fig:instrument}
  \end{center}
\end{figure}

The PAMELA apparatus, shown in Figure \ref{fig:instrument}, is composed by several
sub-detectors: TOF system, anticoincidence system, magnetic spectrometer
with microstrip silicon tracking system, W/Si electromagnetic imaging calorimeter,
shower-tail-catcher scintillator (S4) and neutron detector.
A detailed description of the PAMELA instrument and an overview of the mission can
be found in \cite{pico}.

The instrument has maximum diameter of 102\,cm and height of 120\,cm;
its mass is 470\,kg, the maximum power consumption is 355\,W.
The magnetic spectrometer determines the charge sign and momentum of the incoming particle
through the trajectory reconstruction in the magnetic field; downward-going particles are
identified with the time-of-flight measurement operated by TOF. The final identification
(i.e. antiprotons against electrons etc.) is provided by the combination of the calorimeter
and neutron detector information for kinetic energies above ~1 GeV and by the velocity
measurement (obtained from the trajectory and time-of-flight) at lower energies.
\newline In what follows a synthetic description of the TOF system will be given; further details
on the TOF detectors and electronics can be found in references \cite{tof2008} and
\cite{oste1} respectively.
\newline The TOF system is composed of 6 layers of segmented plastic scintillators, arranged
in three double planes (S1, S2, S3 in Figure \ref{fig:instrument}). The distance between S1
and S2 is around 30\,cm, while the S1-S3 distance is around 77\,cm. Each layer is divided
into several identical paddles (strips), whose number and dimensions vary from layer to layer,
for a total of 24 paddles. For each double plane, the paddles of the upper layer are orthogonal
to those of the lower layer, therefore allowing to get a bidimensional geometrical measurement
of the impact point of charged particles.
The plastic scintillator material, BC-404, manufactured by Bicron company, is characterized by a
rise time of 700\,ps and decay time of 1.8\,ns. Both ends of each scintillator paddle are glued
to an adiabatic UV-transmitting plexiglas light guide. The gluing is obtained with an optical
cement, mod. BC-600 manufactured by Bicron. Each paddle is read-out at each of its two ends by
a photomultiplier tube (PMT) mod. R5900 by Hamamatsu Photonics.
The R5900 is a 10-stage metal package head-on PMT, with rise time of 1.5\,ns, achieving an
amplification of about $4\,\times\,10^6$~at 900\,V. It has a square section of $25.7\,\times\,25.7$\,mm$^2$
~and was chosen for its mechanical robustness, limited size and small weight (25.5\,g).
Since the core of the PAMELA apparatus is a permanent magnet, the PMTs have been shielded from
the residual magnetic field of the spectrometer with a 1\,mm thick $\mu$-metal screen \cite{tof2008}.
\newline The anode pulse of each PMT is converted both in charge (ADC) and time (TDC). The ADC
measurements can be used to determine the Z of the incoming particle. The combined TDC
information of the whole TOF is used to generate the main PAMELA trigger and to measure the
flight time of the incoming particle. The geometry of the TOF planes has been chosen to match
the geometric acceptance of the spectrometer. The standard trigger configuration requires
the coincidence of at least one TDC signal from each of the three TOF double planes.

%*****************************************************************************************
%*****************************************************************************************
%--------------------------------------------
%%%%%%%%%%%%%%%%%%%%%%%%%%%%%%%%%%%%%%%%%%%%%
%%%%%%%%%%%%%%%%%%%%%%%%%%%%%%%%%%%%%%%%%%%%%
\section{GSI beam test}
%%%%%%%%%%%%%%%%%%%%%%%%%%%%%%%%%%%%%%%%%%%%%
%%%%%%%%%%%%%%%%%%%%%%%%%%%%%%%%%%%%%%%%%%%%%
%--------------------------------------------
\label{btest}

\begin{figure}
  \begin{center}
    \mbox{\includegraphics*[scale=0.4]{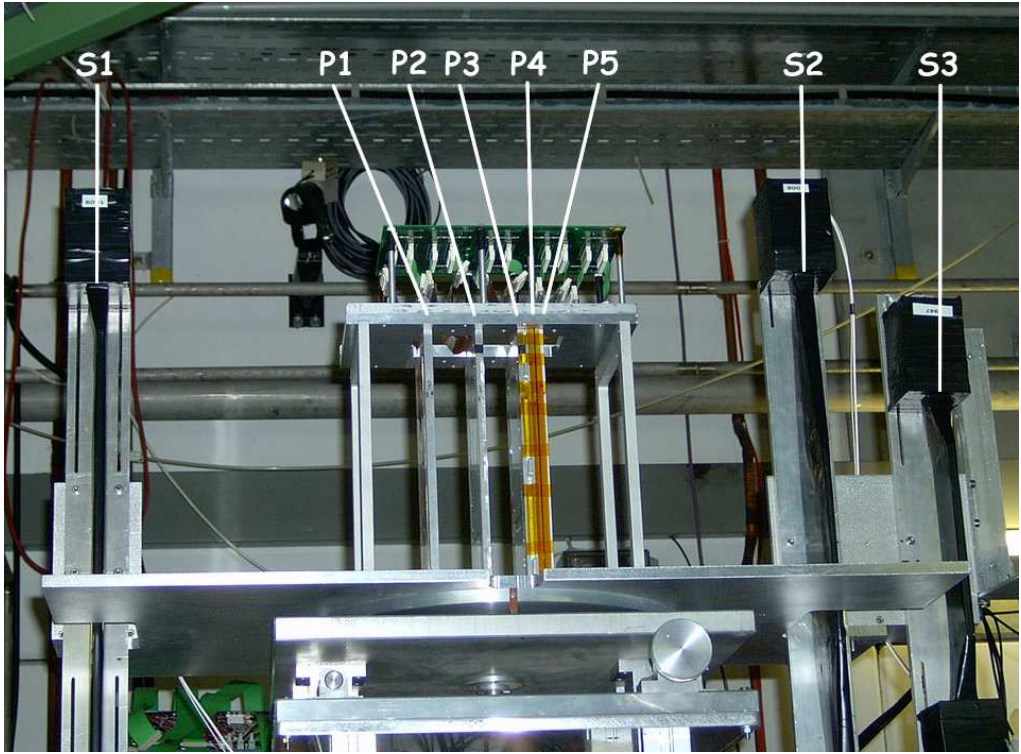}}
    %\hspace{-3mm}\includegraphics*
    %[height=7cm,width=0.5\textwidth,angle=0,clip]{pamelaphoto.eps}}
    \caption{The experimental set-up of PAMELA prototype at the GSI beam
      test.   S1, S2 and S3 are the TOF system scintillators, while P1$\div$P5 are the silicon modules of
      the tracking system.}
    \label{gsi}
  \end{center}
\end{figure}

During its construction phase, PAMELA was tested three times with beams of protons and electrons
at the CERN SPS accelerator, to study the performance of the subdetectors with relativistic
particles. Because of the tight schedule of PAMELA integration, however, it was not possible
to perform a beam test of the flight model with light nuclei before its delivery to Russia
in March 2005.
For this reason such a light-nuclei test  was performed by using  prototypes of the PAMELA
TOF and tracking system in a dedicated mechanical arrangement, on February 2006 at the GSI
(Gesellschaft f$\ddot{u}$r Schwerionenforschung, Darmstadt, Germany) beam accelerator.
Figure \ref{gsi} shows the arrangement of the instrument under test: in this set-up each
TOF plane is formed by just one paddle (indicated as S1, S2, S3 in the picture), while
the prototype of tracking system is composed by 5 Si detector modules (P1 to P5).

The main aim of the test was the determination of the time resolution of the TOF system and of the
charge resolution both for the TOF and for the tracking system for light nuclei. In particular, the
behaviour of these detectors was studied, to determine the amount of variation from linearity
of the corresponding read-out electronics with energy releases due to $Z>1$ particles.
The test results on TOF performance will be presented in the next sections; results for
tracking system will not be discussed here as they will be published elsewhere.\newline
Data were taken during four full days. PAMELA was the secondary beam user for part of this time
(during nights); when main user, it was possible to choose the best available value for beam
intensity (around 1000 particles/spill, with a 4\,s spill duration and a 3\,s interval between
subsequent spills); otherwise, the intensity was much higher (up to $10^8$ particles/spill) thus
resulting in high trigger rate but also in a large fraction of multi-particle events in the detectors.
For some runs, polyethylene or aluminium targets were employed to produce secondaries; besides,
during night hours the presence of biological test-tubes of the main beam user, upstream the
PAMELA prototype, acted as a target. The instrument was normally placed along the beam line,
one meter beyond the target position. For some acquisitions the instrument was moved to a
different position, at the same distance but along a 45$^\circ$ radial line from the target with
respect to the beam line, to record only secondaries scattered at this high angle;
these particles were mostly low-energy protons and He nuclei. Table
\ref{tablegsi} summarizes the different configurations used for the test. Three primary beams
were available: $^{12}$C with kinetic energy of 1200\,MeV/n, $^{12}$C with 200\,MeV/n, $^{50}$Cr with 500\,MeV/n.

The TOF paddles used at GSI are identical to the ones of the corresponding flight-model layer;
the dimensions of the paddles selected for the GSI test are: ($40.8\,\times\,5.5\,\times\,0.7$)\,cm$^3$
for S1, ($18.0\,\times\,7.5\,\times\,0.5$)\,cm$^3$ for S2 and ($18.0\,\times\,5.0\,\times\,0.7$)\,cm$^3$
for S3. The three paddles were arranged with their main (longitudinal) dimension along the
vertical axis. The S1-S3 distance at GSI was around 80\,cm \cite{barba}, while the S1-S2
distance was around 67 cm, greater than in the flight model.
For this beam test the high voltages of the PMTs have been chosen in such a
way to obtain a gain of $1\,\times\,10^6$ which, according to our
calibrations, correspond to HV values varying from 780 V to 820 V.

%*********** [da rimarcare il fatto che la linea anodica
%sarebbe comunque indispensabile dal punto di vista della misura temporale TDC,
%visto che ha il miglior rapporto segnale-rumore rispetto ai dinodi]************

As in the flight model, each paddle is read-out by two PMTs, thus giving a
total of six ADC and six TDC signals. The trigger configuration requires the
coincidence of at least one TDC signal from S1 and S2 paddles.
The read-out electronics employed in this test is the same as for the flight model
\cite{oste1}. Concerning the prototype of tracking system used for the test, each
detector module and its corresponding read-out electronics are identical to the ones
employed in the flight model \cite{straulino} \cite{ricciarini}. To simplify the
whole structure five silicon detector modules have been assembled in a simple aluminum
frame in such a way to keep them aligned. The double-sided silicon sensors
($5.33\,\times\,7.00$)\,cm$^2$ provide two independent impact coordinates on each plane.
The high-resistivity n-type Si bulk is segmented with 1024 read-out microstrips for
each side: p+ strips on the junction side (implantation pitch 25.5\,$\mu$m, read-out
pitch 51\,$\mu$m) and n+ strips on the ohmic side (implantation and read-out pitch 66.5\,$\mu$m).
Junction-side (X-view) strips are orthogonal to ohmic-side (Y-view) ones.

\begin{table}
  \begin{center}
    \caption{\label{tablegsi} BEAMS AVAILABLE AT THE GSI TEST}
    \begin{tabular}{|c|c|c|c|c|}
      \hline \hline {\sf\bf Particle}
      &  {\sf\bf Energy (MeV/n)}
      & {\sf\bf Target} &
             {\sf\bf Angle (deg.)} & {\sf\bf Events }  \\
             \hline $^{12}$C &1200 &    &     0  &  269896 \\
             \hline $^{12}$C & 1200 &$\times$ &  0  & 123194 \\
             \hline $^{12}$C & 200 & &      45&   30378 \\
             \hline $^{12}$C & 200 & $\times$ &  45 & 196139 \\
             \hline $^{50}$Cr & 500 & &      0&   15976 \\
             \hline $^{50}$Cr &  500 & $\times$ &  0 & 173960 \\
             \hline $^{50}$Cr &  500 & $\times$ &  45 & 52241 \\ \hline \hline
    \end{tabular}
  \end{center}
\end{table}

%*****************************************************************************************
%*****************************************************************************************
%--------------------------------------------
%%%%%%%%%%%%%%%%%%%%%%%%%%%%%%%%%%%%%%%%%%%%%
%%%%%%%%%%%%%%%%%%%%%%%%%%%%%%%%%%%%%%%%%%%%%
\section{TOF time resolution and $\beta$ measurements}
%%%%%%%%%%%%%%%%%%%%%%%%%%%%%%%%%%%%%%%%%%%%%
%%%%%%%%%%%%%%%%%%%%%%%%%%%%%%%%%%%%%%%%%%%%%
%--------------------------------------------
\label{TOF_time_resolution}

Two different methods have been used here to measure the time resolution of the
TOF system. The first one combines information from the timing 
measurements of the TDCs and the position measurement of the tracking system;
with this method it is possible to get the intrinsic time resolution of
a TOF paddle.
The second approach takes into account only the measurements of the TOF
itself using the measurements of two paddles, as a result one will get 
the time resolution of the full TOF system.
%For both methods we use raw timing measurement without any amplitude or position
%corrections. Only when we consider measurements with low $Z$
%ions (fragments) we introduce amplitude corrections.

In the first method we determine the intrinsic time resolution of a single
paddle by taking information from both TOF and tracker.
The position of the hit point along the scintillator $x$ is
proportional to the difference of the time measurements
$t_1$ and $t_2$ at the two sides of the scintillator
itself:

\begin{equation}\label{timing_difference}
  x = \frac{v_{eff}(t_1-t_2)}{2} + K  \hspace{.5cm}
\end{equation}

where $v_{eff}$ is the signal velocity inside the scintillator. 

\begin{figure}
  \vspace{0.5cm}
  \begin{center}
    \mbox{\includegraphics*[scale=0.7]{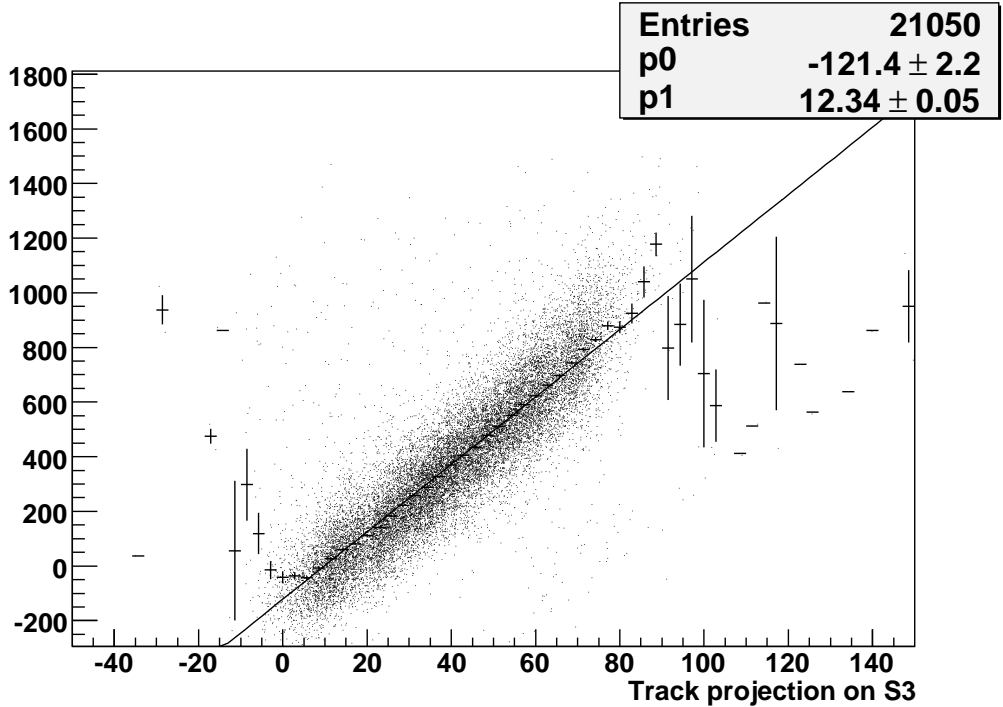}}
    \caption{Scatter plot of the crossing coordinate along paddle S3 as reconstructed 
      through TDC measurement (in units of ps) versus the position measured by the tracking system (in mm). 
      The linear fit and the corresponding bin measurements are superposed. The fit parameter p1 represents $ v_eff$ in cm/ns.
      The points with large error bars are due to poor statistics (regions of the paddle
      reached only by few particles).\label{scat} }
  \end{center}
\end{figure}

If the position of the incident particle along the paddle as determined 
from the timing of the pulses in the two PMTs (in units of picoseconds)
is plotted versus the position as determined by the tracker, we obtain 
the scatter plot shown in figure \ref{scat}. 

\begin{figure}
  \vspace{0.5cm}
  \begin{center}
    %\mbox{\includegraphics*[scale=0.7]{Residui_S3_B_15_02_08_copia.eps}}
    \mbox{\includegraphics*[scale=0.7]{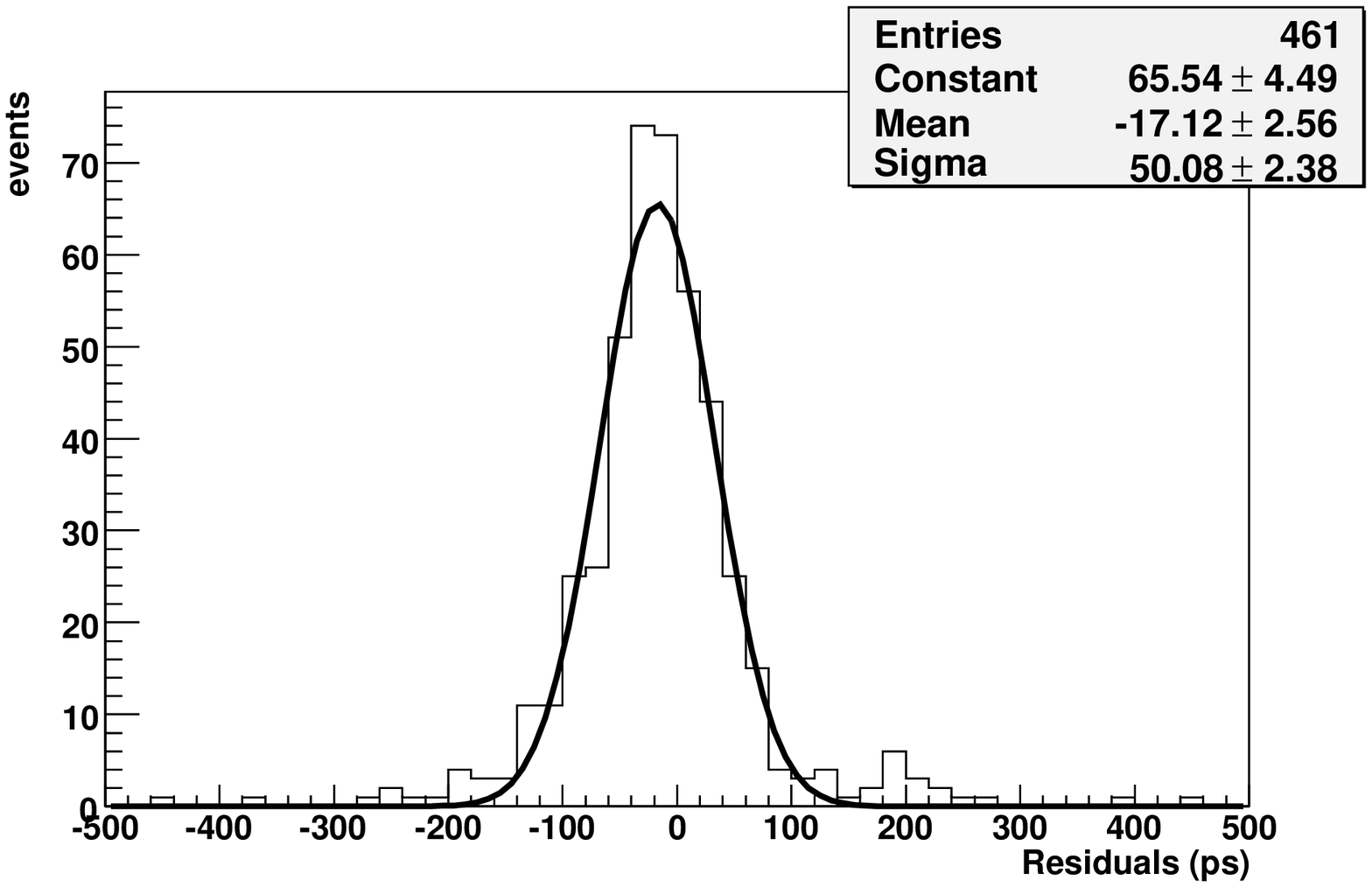}}
    \caption{Residual distribution for Boron sample in the S3 paddle; the standard deviation of this
      distribution is the intrinsic time resolution of this layer in $ps$.\label{resid} }
  \end{center}
\end{figure}

A linear fit to the distribution is shown as well. This fit is used to 
derive the residuals for each event, thus getting finally the distribution 
of timing deviations from the tracker-position which is shown in  figure \ref{resid}. 

Assuming negligible uncertainty in the projected position, the width of a 
Gaussian fitted to this distribution can be taken as the intrinsic time resolution 
of the paddle {\bf $\Delta t_{Si}$}.

%      Scatter plot of the position of the hit points on S3 as reconstructed by
%      the TOF system alone (in units of picoseconds) versus the projection of the 
%      track as reconstructed by the tracking system (in units of mm).
%      The parameter P1 of the fit corresponds to the signal velocity inside the scintillator.
%      The plot refers to the sample of data collected
%      during 45 degree exposition, for which we have such a wide area of projection.

In this way it is possible to evaluate the time resolution of the
single paddle of the TOF system  for each family of nuclei (produced
by fragmentation) from Hydrogen to Carbon, as summarized in table
\ref{tabtimeresZ}. 
As expected, we see an improvement in the time resolution for higher 
charges, since such particles  produce more photons in the scintillator as for a proton of 
equal MeV/nucleon (according to the Bethe-Bloch equation the energy 
release and therefore the number of photons  created in the scintillator
increases with the square of the charge $Z$ of the particle).

\begin{figure}
  \vspace{0.5cm}
  \begin{center}
    %\mbox{\includegraphics*[scale=0.7]{TempoAmpiezza_S3up_15_02_08_copia.eps}}
    \mbox{\includegraphics*[scale=0.7]{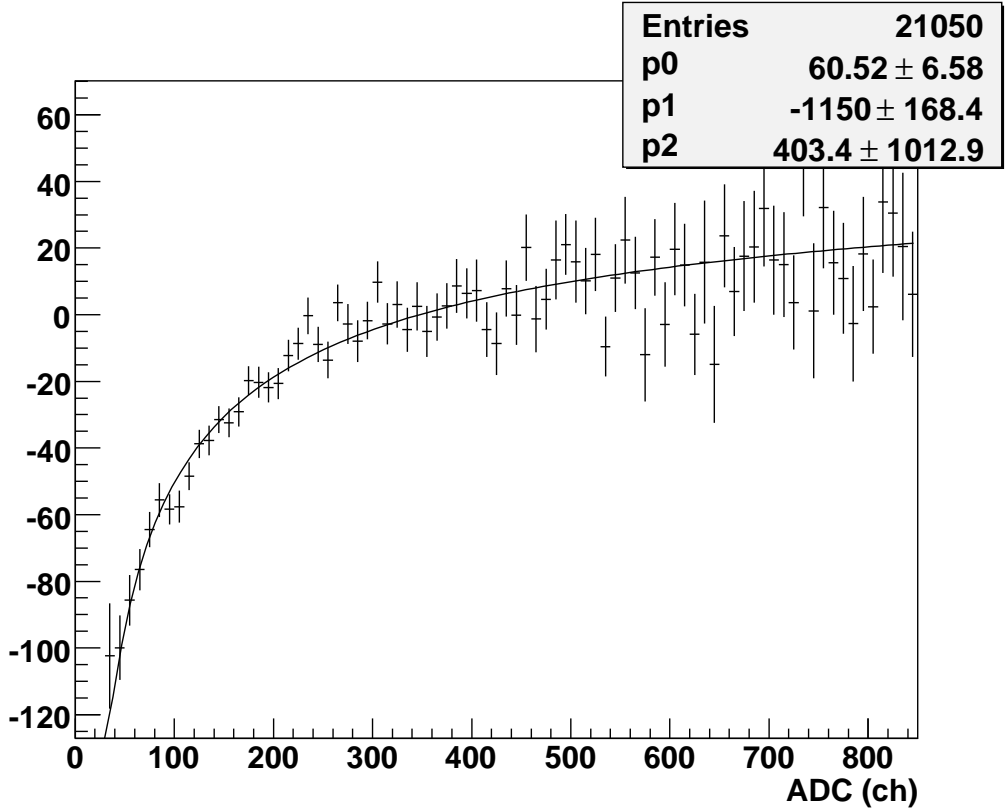}}
    \caption{Differences between the position of the hit point along the paddle S3
      as reconstructed by the TOF alone with respect to the same position reconstructed
      by the tracking system, plotted versus the amplitude (in units of ADC channels) of the signal of one of the two
      PMTs of the paddle. \label{twcorr} }
  \end{center}
\end{figure}

For ions of small $Z$ it is necessary to take into account 
the \emph{Time-Walk\ effect} \cite{braunschweig}.
In order to evaluate the dependence of the time resolution from the 
amplitude of the signal, the quantity to be considered is, for each PMT of a given paddle, the 
residual of each event from a linear fit analogous to that shown
in figure \ref {scat} for the S3 paddle, versus the amplitude of 
the signal of the same PMT.
The points in the resulting plots show a trend which is well fitted (see figure 
\ref{twcorr}) by a typical function  \cite{macro}:

\begin{equation}
  t_{ij} = TDC_{ij} - ( p_0 + \frac{p_1}{\sqrt{ADC_{ij}}} + \frac{p_2}{ADC_{ij}} ) \hspace{.5cm} ,
\end{equation}

where $i$ (=1,2) is the PMT index and $j$ (=1,2,3) is the scintillator paddle index. 
By operating this time-amplitude correction we improved
the time resolution up to Lithium (see last column of table
\ref{tabtimeresZ}).

\begin{table}
  \begin{center}
    \caption{TIME RESOLUTION OF S3 PADDLE FOR DIFFERENT SAMPLES OF
      NUCLEI\label{tabtimeresZ}} \vspace{3mm}
    \begin{tabular}{|c|c|c|}
      \hline \hline {\bf \qquad $Z$ \qquad} & {\bf \qquad $\Delta t$
        (ps)
        \qquad}  &  {\bf $\Delta t$ after Time-Walk correction(ps) } \\
      \hline {1} & 146.5$\pm$0.9 & 117.3$\pm$0.7  \\ \hline {2}
      & 131$\pm$5 & 122$\pm$4  \\ \hline {3} & 118$\pm$4 & 114$\pm$4
      \\ \hline {5} & 50$\pm$2 & 50$\pm$2 \\ \hline {6} &
      46.5$\pm$0.3 & 46.5$\pm$0.3 \\ \hline \hline
    \end{tabular}
  \end{center}
\end{table}

To get the time resolution of the TOF with the second method we use two TOF paddles to 
derive the actual velocity $\beta$ for a particle. 
Using the two paddles $A$ and $B$ we get four TDC measurements, $t_1$ and $t_2$ from paddle $A$,
$t_3$ and $t_4$ from paddle $B$. While the difference of two measurements from a paddle
is proportional to the position of the particle in the paddle (see equation \ref{timing_difference}),
the sum of the two measurements can be taken as the ``mean time" \cite{AMS}. Thus the 
difference of the two sums will be proportional to the particle velocity between the 
two counters.
If we define the ``difference of sums" $DS$ as
 $DS = (t_1 + t_2) - (t_3 + t_4)$, 
 we derive a simple equation between $DS$ and the velocity $\beta$ of the particle
 \cite{carbone}:
\begin{equation}\label{DS}
  DS = K_1 + K_2 \frac{1}{\beta cos\theta} \hspace{.5cm} ,
\end{equation}

where $K_1$ and $K_2$ are two parameters which depend on the experimental
setup, $\theta$ is the zenith angle.
$K_2$ depends solely on known values: $K_2 = \frac{2L}{c}$, where $L$ is the distance between 
the  scintillator paddles and $c$ is the speed of light. 
$K_1$ must be derived from the data itself, since it depends on unknown features of the 
experimental setup like cable lengths. To evaluate $K_1$ we measure $DS$ for
particles of known $\beta$ and invert the equation \ref{DS}. For
this purpose we use, for each of the three types of beam, only data acquired with direct
exposition of the apparatus to the beam (without polyethylene or aluminum target).
The obtained values of $K_1$ for each couple of scintillators are consistent.

% \ref{tableK1}

\begin{figure}
  \vspace{0.5cm}
  \begin{center}
    %\vspace{6cm}
    \mbox{\includegraphics*[scale=0.7]{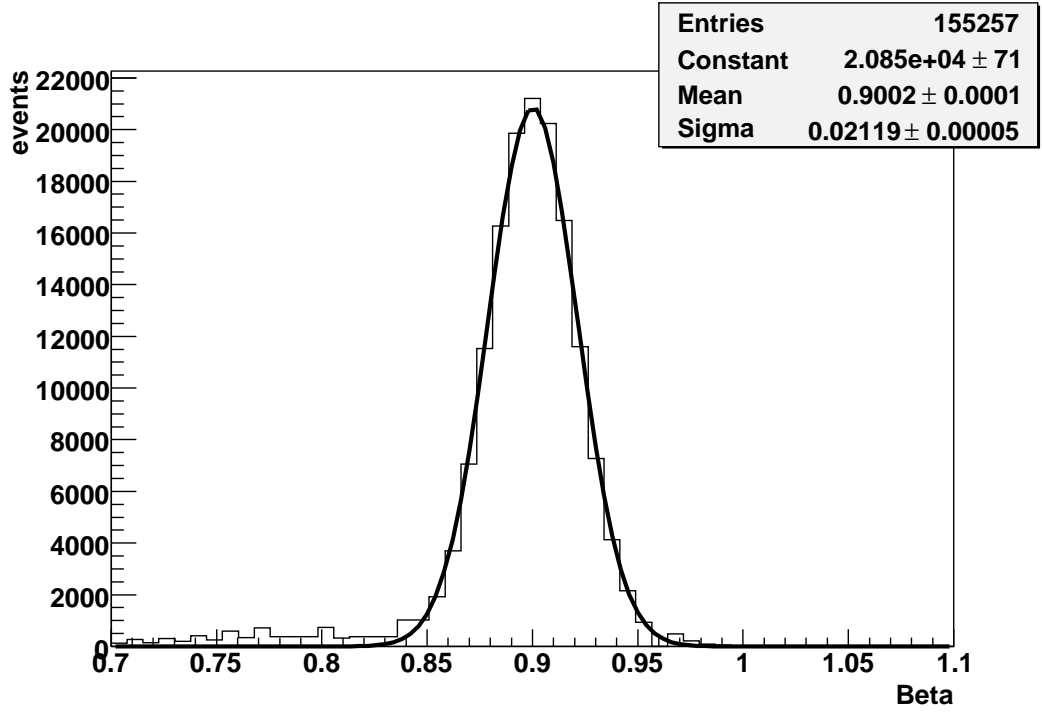}}
    \caption{Distribution of the velocity of particles $\beta$ (in units of speed of
      light) measured between the planes S1 and S2 for a 1200\,MeV/n
      $^{12}$C beam.\label{beta} }
  \end{center}
\end{figure}

With the calculated values of the $K_1$ and $K_2$ constants we can
reconstruct  $\beta$. By exposing the instrument to a monochromatic
beam of particles with fixed kinetic energy (which is the same
sample selected to evaluate $K_1$), the width of the distribution of
the reconstructed $\beta$ (an example is in figure \ref{beta})
determines the time resolution on the measurements of \emph{time of
flight}, according to the simple relation $\Delta t = \Delta \beta
\frac {L}{c \beta^2}$.

Table \ref{tabletimeres} summarizes the results of the
measurements of time resolution of the TOF system  with different
beams and for different combinations of paddles. 

\begin{table}
  \begin{center}
    \caption{\label{tabletimeres} Resolution for time-of-flight and beta measurement}
    \vspace{3mm}
    \begin{tabular}{|c|c|c|c|c|c|}
      \hline \hline {\bf  Ions } & {\bf Paddles }  &  {\bf  Theor.
        $\beta$} &
             {\bf  Measured Mean $\beta$ } & {\bf $\Delta t$(ps)} \\
             \hline {$^{12}$C 1200\,MeV/n}  & S1-S2 & 0.899 & 0.90$\pm$0.02 & 67
             \\ \hline
             & S1-S3 & & 0.902$\pm$0.019 & 62 \\ \hline
                {$^{12}$C 200\,MeV/n} & S1-S2 & 0.568 & 0.568$\pm$0.009 & 61
                \\ \hline
                & S1-S3 & & 0.570$\pm$0.007 & 62 \\ \hline
                   {$^{50}$Cr 500\,MeV/n} & S1-S2 & 0.759 & 0.760$\pm$0.016 & 63
                   \\ \hline
                   & S1-S3 & & 0.760$\pm$0.015 & 68  \\ \hline \hline
    \end{tabular}
  \end{center}
\end{table}

For the determination of the particle velocity it is possible to use any combination 
of planes, like also planes S2 and S3, but being the distance between them of only 13 cm,
the error $\Delta\beta$ will be much larger than for the other two combinations, therefore
it was not used in our analysis.

The measured resolution is consistent with expectations and with tests in laboratory
\cite{oste2}. 
Since the actual $\beta$ for a particle is derived from two paddles, we expect to the 
first order the simple propagation of errors:

\begin{equation}
  \Delta t_{jk} = \sqrt{(\Delta t_{Sj})^2 + (\Delta t_{Sk})^2} \hspace{.5cm} ,
\end{equation}

where $\Delta t_{jk}$ is the resolution of the full TOF using $DS$, and $\Delta t_{Sj}$ and
$\Delta t_{Sk}$ are the intrinsic errors of the paddles derived with the first method.
In the most simple case this will just give a factor $\sqrt{2}$ if the paddles have the 
same intrinsic resolution. Results for C nuclei are in agreement with the results for the 
intrinsic resolution, as one can notice comparing values from tables \ref{tabtimeresZ} 
and \ref{tabletimeres}, being $ \Delta t_{Si} \simeq \Delta t /
\sqrt{2}$ from the DS method.

%---------------------------------------------
%%%%%%%%%%%%%%%%%%%%%%%%%%%%%%%%%%%%%%%%%%%%%
%%%%%%%%%%%%%%%%%%%%%%%%%%%%%%%%%%%%%%%%%%%%%
\section{Charge resolution for light nuclei}
%%%%%%%%%%%%%%%%%%%%%%%%%%%%%%%%%%%%%%%%%%%%%
%%%%%%%%%%%%%%%%%%%%%%%%%%%%%%%%%%%%%%%%%%%%%
%--------------------------------------------
\label{Charge_resolution_for_light_nuclei}

%Due to limitations on total weight and power budget for the TOF system,
%it wasn't possible to use the dynode signal of the PMT so our dynamical range is relatively limited.

To study the charge resolution  of  TOF for light nuclei we tried to have
a data sample widely spread out  in energy so to simulate as well
as possible the situation in flight. Therefore, the full available statistics for C beams has been considered, namely the
$^{12}$C beam at 1200\,MeV/n, with and without target, and the
sample at 200\,MeV/n, with and without target, and recorded at
angles of both 0$^\circ$ and 45$^\circ$.
First step of the analysis was the selection of the data sample to be analyzed. 
The initial data volume was reduced by an amount of 10-15$\%$ eliminating
noisy events or small runs acquired in improper conditions.

%we  eliminated noisy signals in
%the scintillators or wrong TDC values due to improper data unpacking
%for sporadic events. These cuts reduced the sample of about 10-15$\%$.

The particle charge is determined by the energy deposits in any of the
three TOF planes in conjunction with the velocity measurement from
the TOF, that can be derived both by the top and the central and by the top and the bottom
scintillators. The three scintillator layers enable three
independent charge determinations, thus improving significantly
the charge resolution.

\begin{figure}
  \vspace{0.5cm}
  \begin{center}
    \mbox{\includegraphics*[scale=0.7]{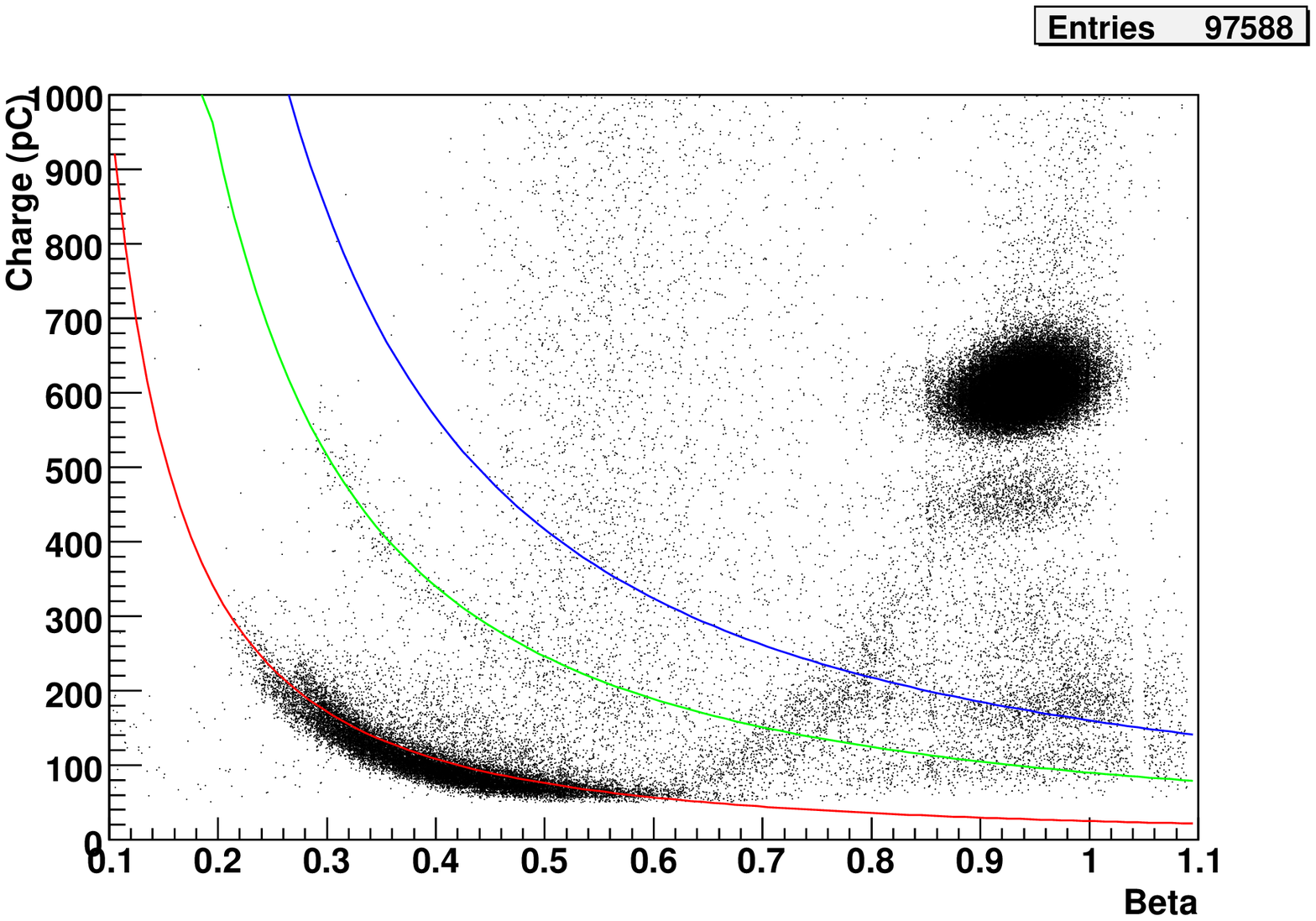}}
    \caption{The energy deposit in S2 as a function of the measured velocity $\beta$.   
      The  particles fall into charge bands. The 3 solid lines identify H, He, Li bands by means of phenomenological functions.
      The dark areas correspond
      to the particles with greater statistics in  our sample. The three darkest areas are: Carbon ions of the beam at 1200\,MeV/n (on the top-right),
       Boron ions (fragment) and low energy protons (recorded at angle of 45$^{\circ}$). The $\beta $ value used in this plot is not corrected for the \emph{Time-Walk\ effect} and this explains the relatively big number of He and Li nuclei in the region with $\beta > 1$. \label{bands}}
  \end{center}
\end{figure}

The measurement of the energy
released inside the scintillator by the passing particle is
proportional to the mean charge deposited, Q, which can be measured by converting the ADC signal (in units of ADC channels)
into charge (in units of $pC$) and correcting this value for
the attenuation of light in the scintillator.
By plotting this charge measurement versus the particle velocity
we saw that points
related to nuclei of different Z fall into different bands (figure \ref{bands});
by fitting these bands it was possible to assign to every Z a mean value of charge deposit at the minimum of ionization.

\begin{figure}
\vspace{0.5cm}
\begin{center}
\mbox{\includegraphics*[scale=0.7]{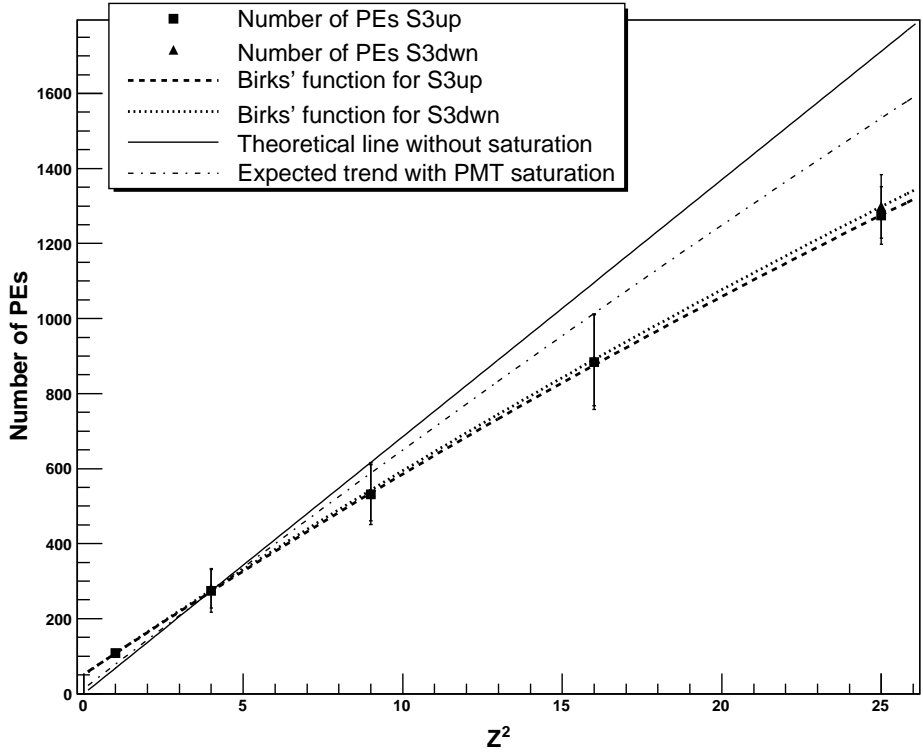}}
 \caption{Loss of linearity for S3 paddle versus $Z^2$ of the incident particle.\label{birks3} }
\end{center}
\end{figure}

The results show that Q increases linearly with $Z^2$ in good approximation for S1 and S2, while for S3 a loss of linearity is observed (see figure \ref{birks3}).
%We saw that the quantity of charge released increased linearly in good
%approximation with $Z^2$ for S1 and S2, but this was not so true for S3.
The behaviour can be justified by looking at the number of photoelectrons (PE) produced in each PMT, which is related to Q through the formula:
%To investigate this behaviour we converted charge information into
%numbers of photoelectrons produced, $PE$, using the formula:

\begin{equation}
 PE = \frac{Q}{e\cdot G} \hspace{.5cm} ,
\end{equation}

where $Q$ is the released charge, $e$ is the charge of the electron
and $G$ is the gain of the PMT.
The  mean number of photoelectrons produced in S3 is
greater than the one produced in S1 and S2 paddles, because S3 is
thicker than S2 (more photons produced) and shorter than S1 (less
attenuation). Apparently, hence, the  charge released in S3 by the
heavier ions covers a region of the dynamical range in which the
system loses linearity.

%The contribution of the PMT to the non linearity of the system was evaluated

%The different behavior of S3 can be explained by considering
%that, for the same particle, due to the different geometry of the three counters,

%In fact, the counter S3 is more thick then S2 and has the same thickness of S3, but is shorter
%and, in our setup, the beam hits the paddles in the center.

To evaluate the contribution of this non-linearity of the PMT output, we observe that,
 with our electronic base, the measured  gain
value of the R5900 PMT's  is constant when the number of photoelectrons is lower than $\sim 700$;
above this value the relation between the two quantities deviates
gradually from linearity 
\footnote{Due to limitations on total weight and power budget for the TOF
system, it was not possible to set-up a separate ADC acquisition chain
exploiting the dynode signal of the PMT or use a different electronic base.}.
This deviation corresponds to
a loss of gain of about 10$\%$ from $\sim 700$ to $\sim 1100$ PE (Be
region) and about 15$\%$ from $\sim 1000$ to $\sim 1500$ PE (B region).

Figure \ref{birks3} shows the comparison between the ``ideal" behaviour of $PE$ versus $Z^2$ (full line, obtained assuming 
that the gain is constant with Z) with the expected dependence taking into account the loss of gain of PMT's (dashed-dotted line) and the actual measurements of $PE$ operated by the two PMTs associated to S3  which are affected by PMT saturation.
Apparently the saturation of the PMTs is not
sufficient to explain the loss of linearity.
The experimental points were
 fitted by a  3-parameters  function (dotted lines):

\begin{equation}
 PE = p_0 +\frac{p_1 Z^2}{1+ p_2 Z^2} \hspace{.5cm} .
\end{equation}

By using this calibration function for a given PMT,
it is possible to associate a value of Z to each particle by measuring the number of
photoelectrons. The particle Z measured by a given paddle can be calculated as the mean between the two independent PMT measurements.

%Once the particle charge, $Z^2$, has been determined
%for each PMT of the paddle, the final $Z$ measured by the plane will
%be the mean of the values obtained by the two PMTs of the paddle.

\begin{figure}
  \vspace{0.5cm}
  \begin{center}
    \mbox{\includegraphics*[scale=0.7]{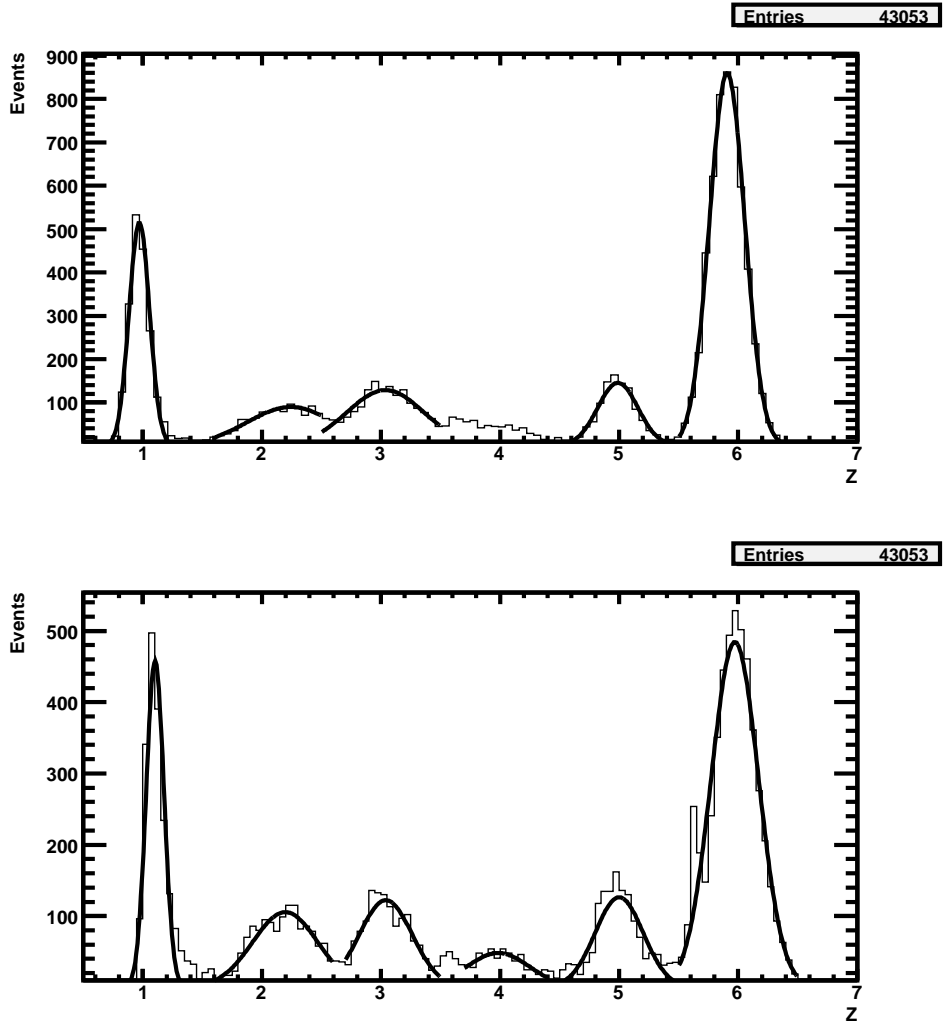}}
    \caption{Charge distribution obtained for particles with different beta and incident angle as measured
      by the TOF (top: paddle S2; bottom: paddle S3). The number of events for $Z=1$ and $Z=6$ are
      divided by a factor 8, in order to make more visible the peaks for
      different values of $Z$. Gaussian fits are superposed on the data. \label{zeta} }
  \end{center}
\end{figure}

The plot of Z distribution for the paddles S2 and S3 is shown in
figure \ref{zeta}, with Gaussian fits superposed on the  data.
%where gaussian fits are superimposed to the data.
The number of events relative to Protons and Carbons are divided by
a factor 8 in order to make more visible the
peaks for different values of $Z$.

Table \ref{tableZTOF} shows the charge resolutions (standard deviations of the Gaussian fits) obtained 
with the previous method for nuclei from H to C and for S2 and S3 paddles.
%The results for charge discrimination for nuclei from H to He are summarized
%in table \ref{tableZTOF}, both for the S2 and the S3 paddle.
The charge uncertainty is less than 0.1 for protons and  0.16 for C (in units of proton charge e).
%We can see that we have a charge uncertainty smaller than 0.1 units of charge for protons and about 0.16 for Carbon.
These results  are extremely satisfactory for PAMELA TOF, since they are of the same order of magnitude of other space missions
that measured nuclei and isotopes in the cosmic radiation, like the ISOMAX balloon-borne mission \cite{isomax}.

\begin{table}
  \begin{center}
    \caption{\label{tableZTOF} CHARGE DISCRIMINATION FOR TOF}
    \vspace{3mm}
    \begin{tabular}{|c|c|c|c|}
      \hline \hline
             {\bf  Nuclei } & {\bf $Z$} & {\bf Paddle } &
             {\bf $\sigma_Z$} \\
             \hline
             H & 1 & S2  & 0.08 \\ \hline
             &  & S3  & 0.07 \\ \hline
             He & 2 & S2  & 0.4 \\ \hline
             &  & S3  & 0.3 \\ \hline
             Li & 3 & S2  & 0.3 \\ \hline
             &  & S3  & 0.2 \\ \hline
             Be & 4 & S2   &  \\ \hline
             &  & S3  & 0.3 \\ \hline
             B & 5 & S2  & 0.17 \\ \hline
             &  & S3  & 0.2 \\ \hline
             C & 6 & S2  & 0.15 \\ \hline
             &  & S3  & 0.17 \\ \hline \hline
    \end{tabular}
  \end{center}
\end{table}

%--------------------------------------------
%%%%%%%%%%%%%%%%%%%%%%%%%%%%%%%%%%%%%%%%%%%%%
%%%%%%%%%%%%%%%%%%%%%%%%%%%%%%%%%%%%%%%%%%%%%
\section*{Conclusion}
%%%%%%%%%%%%%%%%%%%%%%%%%%%%%%%%%%%%%%%%%%%%%
%%%%%%%%%%%%%%%%%%%%%%%%%%%%%%%%%%%%%%%%%%%%%
%--------------------------------------------
\label{Conclusion}

This paper has shown the charge identification capabilities of
PAMELA Time-Of-Flight system, as evaluated during a beam test.
The test was performed at the GSI Laboratory in Darmstadt (Germany),
in February 2006, with a technological copy of the PAMELA TOF and tracking system.

Monochromatic beams of Carbon and Chromium were used for the test,
which lasted 4 days in a 24 hours/day cycle. By means of
polyethylene and aluminum targets,
and positioning the instrument at different angles with respect to the beam axis, it was possible to study the  charge
resolution of the TOF for all light nuclei from Hydrogen to Carbon,
and across a wide energy interval. Results  show that the
PAMELA  Time-Of-Flight reaches very good performance in the
identification of light-nuclei, thanks to the design and quality of
the scintillating paddles.

Furthermore, beam test data were used to estimate the time resolution of the TOF, which resulted in agreement with laboratory tests.

%--------------------------------------------
%%%%%%%%%%%%%%%%%%%%%%%%%%%%%%%%%%%%%%%%%%%%%
%%%%%%%%%%%%%%%%%%%%%%%%%%%%%%%%%%%%%%%%%%%%%
\section*{Acknowledgments}
%%%%%%%%%%%%%%%%%%%%%%%%%%%%%%%%%%%%%%%%%%%%%
%%%%%%%%%%%%%%%%%%%%%%%%%%%%%%%%%%%%%%%%%%%%%
%--------------------------------------------
\label{Acknowledgments}

We acknowledge the staff working at GSI, and especially dr. Dieter
Schardt, for the excellent professionalism and friendly
collaboration they offered us during our work in the laboratory.
We would like also to thank the followings technicians of the INFN structure
 of Naples for their valuable contribution to the realization of the apparatus:
P. Parascandolo, G. Passeggio, G. Pontoriere, E. Vanzanella.

% References

%If we fit the track position as a function of the time difference in the paddle,
%and hit by hit, we evaluate the residual between the time difference obtained from the interpolated function and the one provided by the TOF,
%we obtain the residual distribution in  figure \ref{resid}.

%Since is derived from two paddles, which should degrade the resolution by a factor of   $\sqrt{2}$
% The first step in the time resolution evaluation is the
% determination of the particle velocity $\beta$. 


\begin{thebibliography}{00}
%\bibitem{maurin} D. Maurin et al., ApJ, 2001,  555, p. 585.
%\bibitem{prishchep} V. S.
%Prishchep \& V. S. Ptuskin, Ap\&SS, 1975, 32,  p. 265.
%\bibitem{ginzburg} V. L.
%Ginzburg \& V. S. Ptuskin, 1976, Rev. Mod. Phys, 48,  p. 161.
%\bibitem{ptuskin}
%V. S. Ptuskin \& A. Soutoul, 1998, A\&A, 337,  p. 859.
%\bibitem{moskalenko} I. V. Moskalenko  \& A. W. Strong, 2000, Ap\&SS,
%272,  p. 247.
%\bibitem{streit}R. E. Streitmatter  \& S. A.
%Stephens, 2001, Adv. Space Res., 27,  p. 743.
%\bibitem{bere} V. S.
%Berezinskii, 1990, Astrophysics of Cosmic Rays, ed. V. L. Ginzburg
%(Amsterdam: North-Holland).
%\bibitem{stephens} S. A. Stephens \& R. E.
%Streitmatter, 1998, ApJ, 505,  p. 266.
%\bibitem{simon} A. Castellina, F. Donato, Astropart. Phys., 2005,  24, p. 146.
%\bibitem{moskalenko2} A. W. Strong, I. V. Moskalenko, V. S. Ptuskin, Annu. Rev. Nucl. Part. Sci., 2007, 57, p. 285.
%
%\bibitem{isomax} The Astrophysical Journal, 611, pag. 892--905, 2004.
%
\bibitem{pico} P. Picozza et al., Astropart. Phys, 2007, 27, p. 296.
\bibitem{tof2008} G. Barbarino et al., NIM A, 2008, 584, p. 319.
\bibitem{oste1} G. Osteria et al., NIM A, 2004, 518, p. 161.
\bibitem{barba} G. Barbarino et al., Nuclear Physics B,  2003, 125, p. 298.
\bibitem{straulino} S. Straulino et al., NIM A, 2004, 530, p. 168.
\bibitem{ricciarini} S. Ricciarini et al., NIM A, 2007, 582, p. 892.
\bibitem{braunschweig} W. Braunschweig et al., NIM 134, 1976, p. 261.
\bibitem{macro} M. Ambrosio et al., NIM A, 2002, 486, p. 663.
\bibitem{AMS} D. Alvisi et al., NIM A, 1999, 437, Issues 2-3, p. 212.
\bibitem{carbone} R. Carbone et al., NIM A, 2008, 588, Issues 1-2, p. 235.
%\bibitem{bess} Y. Shikaze et al., NIM A, 2000, 455, Issue 3, p.596.
\bibitem{oste2} G. Osteria et al., NIM A, 2004, 535, p. 152.
\bibitem{isomax} G. A. de Nolfo et al., The Astrophysical Journal, 2004, 611, p. 892.
\end{thebibliography}
\end{document}